\documentclass[11pt]{article}
\usepackage{amssymb}
\usepackage{epsfig}
 \input prepictex
 \input pictex
 \input postpictex

\let\tsection\section
\renewcommand{\section}{\setcounter{equation}{0}\tsection}

\begin{document}
%\today
\begin{center}

THE ASYMMETRIC EXCLUSION PROCESS AND BROWNIAN EXCURSIONS
\vskip10pt

B. Derrida\footnote{Laboratoire de Physique Statistique,
Ecole Normale Sup\'erieure, 24 rue Lhomond, 75005 Paris, France;
emails derrida@lps.ens.fr., enaud@lps.ens.fr},
C. Enaud${}^{1}$, and
J. L. Lebowitz\footnote{Department of Mathematics and Physics,
Rutgers University, New Brunswick, NJ 08903; email
lebowitz@math.rutgers.edu.}

\end{center}

%\today
\vskip20pt

\vskip20pt
\noindent {\bf Abstract}

 We consider the totally asymmetric exclusion process (TASEP)
in one dimension
 in its maximal current phase. We show, by an exact calculation,
 that the non-Gaussian part of the  fluctuations of density can be described
in terms of the statistical properties of a Brownian excursion.
   Numerical simulations
indicate that the description in terms of a Brownian excursion
remains valid
 for more general one dimensional
driven systems in their maximal current phase.

\vskip.5truein
\noindent  Dedicated to Gianni Jona-Lasinio, who was and continues to
be a pioneer in this field, on the occasion of his seventieth birthday.
\newpage

\section{Introduction\label{introduction}}

Exclusion processes \cite{spohn,ZS,Ligg2} with open boundaries have
attracted much attention as simple models of an open non-equilibrium
system in contact with two reservoirs having different chemical
potentials \cite{krug,DDM,SD}. Despite their simplicity, these models
exhibit properties believed to be characteristic of realistic
non-equilibrium systems, such as long range correlations
\cite{HS1,DE,DLS1,DLSasep,DLSasep2} and phase transitions in one
dimension \cite{krug,DDM,SD}.

A number of exact results have been obtained for the one dimensional
exclusion process with open boundaries, using the fact that the
weights of the microscopic configurations in the stationary state can
be calculated exactly \cite{DDM,SD,DEHP,Sandow,MS,ER,Sas,BECE}.  The
goal of the present paper is to show that the density fluctuations of
the totally asymmetric exclusion process in its stationary maximal
current phase can be expressed in terms the statistical properties of
a Brownian excursion.

A Brownian excursion \cite{Pitman} is a Brownian path $Y(x)$
conditioned to remain positive, i.e. such that $Y(x) >0$ for $0< x<1$,
with the boundary condition $Y(0)=Y(1)=0$.  In the following we will
consider the Brownian excursion $Y(x)$ normalized in such a way that
the probability density of being at heights $y_1, y_2,..  y_k$ at
times $0 < x_1 < x_2 < ... x_k < 1$ (i.e. that $Y(x_1)=
y_1,... Y(x_k)=y_k$) is given by
\begin{equation}
 {\rm  Prob} (y_1,...y_{k };x_1,...x_k)=
 {1  \over \sqrt{\pi } } \  h(y_1;x_1)
\ h(y_k; 1- x_k)
\prod_{p=1}^{k-1} \ g(y_p,y_{p+1};x_{p+1} -x_p), 
\label{pro}
\end{equation}
where the functions $h$ and $g$ are defined by
\begin{equation}
 h(y;x) = {2 y \over x^{3/2}} \  e^{-{y^2 \over x}} 
\label{hdef}
\end{equation}
and
\begin{equation}
 g(y,y';x) = {1  \over \sqrt{\pi x} } \  \left[ e^{-{(y-y')^2 \over x}}
-  e^{-{(y+y')^2 \over x}} \right] .
\label{gdef}
\end{equation}
One can check that (\ref{pro}) is normalized, i.e.\ that
$$ \int_0^\infty dy_1 ...\int_0^\infty dy_k  \   {\rm Prob}
(y_1,...y_{k };x_1,...x_k) =1, $$ 
using the identities 
$${1  \over \sqrt{\pi } } \int_0^\infty dy  \ h(y;x) \ h(y;1-x) =  1, $$
$$\int_0^\infty dy'  \ h(y';x) \ g(y',y;x') =  h(y;x+x'), $$
$$ \int_0^\infty dy''  \ g(y,y'';x) \ g(y'',y';x') =  g(y,y';x+x'). $$

The one dimensional totally asymmetric simple exclusion process
(TASEP) with open boundary conditions is defined as follows: each site
$i$ (with $1 \leq i \leq L$) of a one dimensional lattice of $L$ sites
is either occupied by a single particle or empty, and the system
evolves according to the following continuous time dynamics: if a
particle is present on site $i$ (for $ 1 \leq i \leq L-1$), it hops at
rate $1$ to site $i+1$ if this site is empty. At the left boundary,
site $1$ is filled at rate $\alpha$ by a particle if it is empty.  At
the right boundary, if a particle is present at site $L$, it is
removed at rate $\beta$. Each microscopic configuration can be
described by a set of $L$ binary variables $\tau_i$, the occupation
numbers ($\tau_i=1$ if site $i$ is occupied and $\tau_i=0$ if it is
empty).  When $\alpha$ and $\beta$ lie in the interval $(0,1)$, the
case we shall be concerned with here, the input and exit rates
$\alpha$ and $\beta$ can be thought as resulting from the system being
in contact with a left and and a right reservoir at densities $\alpha$
and $1-\beta$ respectively \cite{DLSasep2}.

In the steady state, which is unique for such a Markov process, one
can try to determine correlation functions, which we will denote $
\langle \tau_{i_1} \tau_{i_2} ... \tau_{i_k} \rangle $.  One can also
divide the $L$ sites into $k$ boxes of $L_1,L_2,... L_k$ sites and try
to determine the probability that in the steady state $N_1$ particles
are present in the first box, $N_2$ in the second box,...  $N_k$ in
the $k$th box.

In the present paper we are going to show that in the maximal current phase
 \cite{krug,DEHP,PS} of the stationary TASEP, corresponding to 
\begin{equation}
\alpha > {1 \over 2}, \ \ \  {\rm and} \ \ \  \beta > {1 \over 2},
\label{max}
\end{equation}
the correlation functions of the occupation numbers $\tau_i$ in the
steady state are given for large $L$ by
\begin{equation}
 \left\langle \left( \tau_{i_1} -{1 \over 2} \right) \left( \tau_{i_2}
-{1 \over 2} \right) ... \left( \tau_{i_k} -{1 \over 2} \right)
\right\rangle \simeq {1 \over 2^k L^{k \over 2}} \ {d^k \over d x_1
dx_2 ... dx_k } \ \overline{ Y(x_1) ... Y(x_k) },
\label{cor}
\end{equation}
where $i_1 <
i_2, ... < i_k$ and the right hand side is to be evaluated at  $x_p
= i_p/L$.  The averages are taken with respect to (\ref{pro}).

Our second result concerns the fluctuations of the numbers
$N_1,... N_k$ of particles in boxes of length $L_1= L (x_1 - x_0) ,
L_2= L(x_2-x_1),...  L_k= L(x_k -x_{k-1}) $ with $ x_0 = 0 < x_1 <
.... < x_{k-1} < x_k=1 $.  Define $\mu_p$ to be the rescaled
fluctuations of the number of particles, in box $p$,
\begin{equation}
\mu_p= {N_p - L(x_p - x_{p-1}) /2 \over \sqrt{L}}.
\label{mup}
\end{equation}
We are going to show that their probability density
$Q(\mu_1,... \mu_k; x_1,... x_{k-1}) $ is given for large $L$ by
\begin{eqnarray}\label{Qmu}
&& Q(\mu_1,...\mu_{k }; x_1,... x_{k-1})= \int_0^\infty dy_1
 ...\int_0^\infty dy_{k-1 } \times \\ \nonumber && {\rm
 Prob}(y_1,...y_{k-1 };x_1,...x_{k-1}) \prod_{i=1}^k {2 \over
 \sqrt{\pi (x_i- x_{i-1})} } \exp \left[- (2 \mu_i+ y_{i-1}- y_i)^2
 \over x_i - x_{i-1} \right]
\end{eqnarray}
where  $y_0=y_k=0, x_0=0$, $x_k=1$

As we shall see in section 2 the product in the integrand of
(\ref{Qmu}) is just the conditional probability of $\mu_1,...,\mu_k$
given the values of the $Y$ process, $Y(x_1)=y_1,...Y(x_{k-1}) = y_{k-1}$.
This conditional probability is just a product of Gaussians with
means $(y_i- y_{i-1} )/2$ and variances $(x_i - x_{i-1})/8$.  Since
(\ref{Qmu}) is valid for arbitrary number and sizes of boxes it is equivalent
to the statement that the ``fluctuation field'' of the particle
density $\rho(x)$ in the maximum current phase can be written as a sum of two
independent processes,
\begin{equation}
\sqrt L \left[\rho(x) - {1 \over 2} \right] \simeq {1 \over
2}\left[\dot B(x) + \dot Y(x)\right].
\label{BM+BE}
\end{equation}
Here $\rho(x)$ is the empirical density at $x$, defined by 
$$
\sqrt L \int_{x_{p-1}}^{x_p} \left(\rho(x) - {1 \over 2} \right)dx
\simeq {1 \over \sqrt L} \sum_{i = x_{p-1}L}^{x_p L}\left(\tau_i - {1
\over 2}\right) = \mu_p,
$$
while $\dot Y(x)$ is the (generalized) derivative of the Brownian
excursion described by (\ref{pro}), $\dot B(x)$ is a white noise, the
derivative of a Brownian path, normalized so that
\begin{equation}
\label{1.9}
 \overline{[B(x)-B(x')]^2}  = {1 \over 2} |x-x'|, \end{equation}
and $B$ and $Y$ are independent.

{}For the integrated fluctuation $r(x,x')$ of the density in the
macroscopic segment $(x, x^\prime)$, 
\begin{equation}
 r(x,x') = \int_x^{x'} dy \left(\rho(y) - {1 \over 2 } \right) \simeq
{ B(x') + Y(x') - B(x) - Y(x) \over 2 \sqrt L }  ,
\label{rxxp}
\end{equation}
one has
\begin{equation}
\overline { r(x,x') ^2 } \simeq { 
\overline{[B(x)-B(x')]^2}
+\overline{[Y(x)-Y(x')]^2} 
\over 4 L}  \ .
\label{r2}
\end{equation}
If one considers now the fluctuation in a very small segment away from
the end points, then $Y(x^\prime) - Y(x)$ behaves like $B(x^\prime) -
B(x)$ and so  
$$\overline { r(x,x') ^2 } \simeq {|x' - x| \over 4  L } \ \ \  \ \ \ {\rm
for} \ |x-x'| \to 0, $$
indicating that locally the measure is Bernoulli.  On the other hand
if one considers the particle number fluctuation  in the whole system 
$$\overline{ r(0,1) ^2 } \simeq {1 \over 8 L}, $$ which means that the
fluctuation of the total number of particles is one half of what it
would be for a Bernoulli measure at density $1/2$ \cite{DE}.

One can check that (\ref{cor}) extends to arbitrary correlations what
was already known for the one-point and two-point functions (equations
(47) and (52) of \cite{DE}) when $\alpha=\beta=1$. Also (\ref{Qmu})
extends to an arbitrary number of boxes the result (6.15) of
\cite{DLSasep2} valid for a single box.

Our derivation presented in section 2 is a generalization of the
method used in \cite{DLSasep2}.  Numerical simulations reported in
section 3 indicate that the description in terms of a Brownian
excursion remains valid in the maximal current phase of other models
for which the steady state is not known exactly.

\section{ Derivation of (\ref{cor}) and (\ref{Qmu}) }
\subsection{The matrix method}
{}For the steady state of the open TASEP described in
Section~\ref{introduction}, the probability $P(\{\tau_i\})$ of any
microscopic configuration $\{\tau_i\}$ can be written as \cite{DEHP} %
\begin{equation} P(\{\tau_i\}) = { \langle W | \prod_{i=1}^N [ D
\tau_i + E(1-\tau_i) ] | V \rangle \over Z_L} ,
\label{matrix}
 \end{equation}
where the normalization factor $Z_L$ is given by
 \begin{equation}
Z_L = \langle W | (D+E)^L | V \rangle ,
\label{normalisation}
 \end{equation}
and the matrices $D$, $E$ and the vectors $\langle W|$, $|V\rangle$
satisfy the relations, 
 \begin{eqnarray}
     DE  &=&  D + E  \label{DE}\;,\label{alg1}\\
  \beta D |V\rangle &=&  |V\rangle\;,\label{alg2}\\
   \langle W|\alpha E  &=&  \langle W|\;\label{alg3}.
  \end{eqnarray}

 {}From the algebra (\ref{matrix})--(\ref{alg3}) all equal time steady state
properties can  be calculated.  {}For example the average
occupation $\langle \tau_i \rangle $ of site $i$ is given by
 \begin{equation}
\langle \tau_i \rangle  
  = {\langle W | (D+E)^{i-1} D (D+E)^{L-i} | V \rangle \over Z_L},
\label{taui}
 \end{equation}
and the two point function is, for $i < j$,
 \begin{equation}
\langle \tau_i \tau_j\rangle  
 = {\langle W | (D+E)^{i-1}  \ D \  (D+E)^{j-i-1} \  D \  (D+E)^{L-j}
 | V \rangle  
   \over Z_L}.
\label{tauitauj}
 \end{equation}

The probability of finding $N_1,..N_k$ particles in subsystems of lengths 
$L_1,...L_k$ can be written as
\begin{equation}
q_{L_1,L_2... L_k}(N_1,..N_k)={\langle W | X_{L_1}(N_1) \
   X_{L_2}(N_2) ...  \  X_{L_k}(N_k) | V \rangle 
   \over Z_L}
\label{PN}
\end{equation}
where $X_{L}(N)$ is the sum over all the products of $L$ matrices
containing exactly 
$N$ matrices $D$ and $L-N$ matrices $E$. 
This can be written as
\begin{equation}
 X_{L}(N) = \int_0^1 d\theta  \ \left( D e^{2 i \pi \theta} + E
 \right)^L \ e^{- 2 i \pi  N \theta} 
\label{XLN}
\end{equation}
The algebraic rules (\ref{alg1}-\ref{alg3}) allow one  to calculate all the
matrix elements appearing in
(\ref{matrix},\ref{normalisation},\ref{taui}--\ref{PN})
without using any explicit representation of the matrices $D$ and $E$
or of the vectors  $\langle W|$ and
 $|V\rangle$.  Working with a particular representation might be
convenient but of course the steady state properties,
such as correlation functions and current,  do not depend on the
particular representation used.

To derive the expressions (\ref{cor},\ref{Qmu}) we find it convenient
 to use a particular 
 representation of  (\ref{alg1}-\ref{alg3})
(which was already   used in section 6.3 of \cite{DLSasep2}):
\begin{eqnarray}
\label{Drep}
  D &=& \sum_{n=1}^\infty | n \rangle \langle n | + |n \rangle \langle n+1|,
   \\
  \label{Erep}
  E &=& \sum_{n=1}^\infty | n \rangle \langle n | + |n+1 \rangle \langle n|,
\end{eqnarray}
 where the vectors $|1 \rangle, |2 \rangle, ...|n \rangle ...$ form an
orthonormal basis of an infinite dimensional space (with $\langle n | m
\rangle = \delta_{n,m}$).  In this basis, the vectors $|V \rangle$ and
$\langle W|$ satisfying (\ref{alg2}),(\ref{alg3}) are given
 by
\begin{eqnarray}
\label{Vrep}
  |V \rangle 
  = \sum_{n=1}^\infty  \left( 1 - \beta \over \beta \right)^n | n \rangle
,
  \\
 \label{Wrep}
 \langle W | 
  = \sum_{n=1}^\infty  \left( 1 - \alpha \over \alpha \right)^n \langle  n  |.
\end{eqnarray}
In the following we will assume that
\begin{equation}
1 < \alpha + \beta \ \ \ , \ \ \ \alpha <1, \ \ \ {\rm and } \ \ \ \beta <1.
\label{condition}
\end{equation}
so that $ \langle W| V \rangle$ and all the matrix elements $ \langle
W| X_1 X_2 ... | V \rangle$ are finite and positive when the matrices
$X_1,X_2..$ are polynomials of matrices $D$ and $E$ with positive
coefficients.  This condition is not the same as the condition
(\ref{max}) of being in the maximal current phase. We will see later
(\ref{maxicur}) how condition (\ref{max}) appears.

Note that as long as $L$ is finite, all the matrix elements are
rational functions of $\alpha$ and $\beta$ and so all the expressions
derived assuming (\ref{condition}) could be analytically continued to
the whole range of values of $\alpha$ and $\beta$.

\subsection{The sum over walks}
Let us introduce the set of discrete walks $w$ of $L$ steps which, at
each step, either increase by one unit, decrease by one unit, or stay
constant, with the constraint that they remain positive.  Each such
walk $w$ can be described by a sequence of $L+1$ integers $\{n_{i} (w)
\}$ satisfying for all $i = 0,1,...,L$
$$n_i > 0 \ \ \ \ {\rm and} \ \ \ \ |n_i - n_{i-1}| \leq 1$$
To each such walk $w$, one associates a weight $\Omega(w)$ defined
by
\begin{equation}
\Omega(w)= 
  \left( 1 - \alpha \over \alpha \right)^{n_0} 
 \left( 1 - \beta \over \beta \right)^{n_L} 
\prod_{i=1}^{L} v \left(n_{i-1}, n_{i} \right)
\label{Omega}
\end{equation}
where $v(n,n')$ is given by
$$v(n,n') = \cases{ 2 \ \ \ \ {\rm if} \ \  |n-n'|=0 \cr
 1 \ \ \ \ {\rm if} \ \  |n-n'|=1 \cr
 0 \ \ \ \ {\rm if} \ \  |n-n'|>1 } $$
It is easy to check from (\ref{Drep},\ref{Erep}) that for $n \geq 1$
and $n' \geq 1$, one has 
$$ v(n,n') =  \langle n |  D+E | n' \rangle $$
and it follows that
\begin{equation}
 \langle W | (D+E)^L | V \rangle = \sum_{w}  \Omega(w)  .
\label{norma}
\end{equation}
These weights define a measure $\nu(w)$ on the walks $w$
\begin{equation}
\label{nu}
\nu(w)= { \Omega(w) \over \sum_{w'} \Omega(w') } \ \ \  .
\end{equation}
It follows from (\ref{Drep},\ref{Erep})  that
$$\langle n| D| n' \rangle  = {(1 + n'-n)   \langle n| D +E | n'
\rangle \over 2}, $$
which combined with  (\ref{tauitauj}) yields, for $i < j$, 
\begin{equation}
 \left\langle \tau_{i} \tau_{j} \right\rangle =  {1 \over 4 }  \sum_{w}
\nu(w)  \left[ 1+ n_{i}(w) - n_{i-1}(w)   \right]
 \left[ {1+ n_{j}(w) - n_{j-1}}(w)   \right] .
\end{equation}
More generally, for $i_1 < i_2 < ... < i_k$, 
\begin{equation}
 \left\langle \tau_{i_1}.... \tau_{i_k} \right\rangle = {1 \over 2^k}
\sum_{w} 
\nu(w)  \left[ 1+ n_{i_1}- n_{i_1-1}  \right] ....
 \left[ {1+ n_{i_k}- n_{i_{k}-1}}  \right]  ,
\label{eq1}
\end{equation}
(where to avoid heavy notation we have not repeated the $w$ dependence of
all the $n_i$'s).
This can be rewritten as
\begin{equation}
 \left\langle \left(\tau_{i_1}-{1 \over 2} \right)....
\left(\tau_{i_k}-{1 \over 2} \right) \right\rangle = 
 {1 \over  2^k }
 \sum_{w}
\nu(w)  \left[  n_{i_1}- n_{i_1-1}  \right] ....
 \left[ { n_{i_k}- n_{i_{k}-1}}  \right], 
\label{eq2}
\end{equation}
which is the exact finite $L$ version of (\ref{cor}).

The expressions of  $ \langle n | (D+E)^L | n' \rangle$ and of
$\langle n| X_{L}(M) | n' \rangle$ 
defined in (\ref{XLN}) are known (see equations (6.24) and (6.65) of
\cite{DLSasep2}  which had been derived  
 by recursions over $L$ in \cite{Muk,Mal}).
 \begin{equation}
  \langle n | (D+E)^L | n' \rangle = { (2L)! \over (L+n-n')! \  (L+n'-n)! }
- { (2L)! \over (L+n+n')!  \ (L-n'-n)! }, 
\label{DplusE}
 \end{equation}
\begin{eqnarray}
  \langle n| X_L(N) | n' \rangle =
  { (L!)^2 \over (N)! \  (L-N)! \  (N+n-n')! \  (L-N-n+n')!}
\nonumber \\
- { (L!)^2 \over (N+n)! \  (L-N-n)! \  (N-n')! \   (L-N+n')!}, 
\label{XLNexp}
 \end{eqnarray}
where any negative factorial is defined to be infinite (i.e. the matrix
elements are non-zero only when $N - L \leq n^\prime - n \leq N \leq L$).  

Let 
\begin{equation}
F_{L,N}(n,n') = { \langle n| X_L(N) | n' \rangle \over \langle n |
(D+E)^L | n' \rangle } .
\label{F}
 \end{equation}
The probability $q_{L_1,L_2... L_k}(N_1,..N_k)$  that there  are $N_1$
particles in 
the first $L_1$ sites, $N_2$ in the next $L_2$ sites, etc., is then given by
\begin{equation}
\label{qL}
q_{L_1,L_2... L_k}(N_1,..N_k)= 
  \sum_{w}
\nu(w)  \prod_{i=1}^k  F_{L_i,N_i}(n_{M_{i-1}}(w),n_{M_i}(w))
\end{equation}
where $M_0=0$ and $M_{i}=M_{i-1} + L_i$.
This is the exact finite $L$ version of (\ref{Qmu}).  It shows clearly
that given $w$ the $\{N_i\}$ are independent random variables.  

\subsection{Derivation of (\ref{cor})}

Let us first
  evaluate for large $L$  the normalization factor (\ref{norma})
\begin{equation}
\label{sumw}
  \sum_{w}
 \Omega(w) = \sum_{n=1}^\infty
  \sum_{n'=1}^\infty
 \left( 1 - \alpha \over \alpha \right)^{n}
 \left( 1 - \beta \over \beta \right)^{n'}
\langle n | (D+E)^L |n' \rangle
\end{equation}
Using (\ref{DplusE}) (or its large $L$ behavior easily obtained by the
Stirling formula) one can show that if
\begin{equation}
\alpha > {1 \over 2}, \ \ \ \ {\rm and}  \ \ \ \ \beta > {1 \over 2}, 
\label{maxicur}
\end{equation}
then the walks which dominate the sum for large $L$ are those which
have 
both $n_0(w)$ and $n_L(w)$ of order $1$  and which remain at distances
of order $L^{1/2}$ from the origin.
%(i.e. such that $n_i(w) \sim L^{1/2}$ for all $i$).
Condition (\ref{maxicur}) corresponds to the maximal current phase
\cite{DEHP,PS}.
It is  more restrictive than (\ref{condition}) which assures only
that $\langle W| V \rangle $ is finite.
In the range where (\ref{condition}) is satisfied but 
(\ref{maxicur}) is not, the walks which dominate (\ref{sumw}) are
walks such that either $n_0(w)$ or $n_L(w)$ is of order $L$, and where
the Brownian excursion picture does not apply.

%In fact, when (\ref{maxicur}) is satisfied then, as can be seen from
%the equations below, 
%$n_0$ and $n_L$ become independent for $L \to \infty$ with
%$$
%{\rm Prob}(n_0) = ({1-2 \alpha \over \alpha})^2 n_0 ({1 - \alpha
%\over \alpha})^{n_0} 
%$$
%$$
%{\rm Prob}(n_L) = ({1 - 2 \beta \over \beta})^2 n_L ({1 - \beta \over
%\beta})^{n_L}.
%$$
The large $L$ expressions 
of  matrix elements of the form (\ref{DplusE}) can be easily obtained
using Stirling formula and one gets, if 
and $n$ and $n'$ are of order $\sqrt{L}$, 
\begin{equation}
\langle n | (D+E)^{L} | n' \rangle 
 \simeq { 4^{L} \over \sqrt{\pi L} }
\left[ \exp \left(-(n-n')^2 \over L \right)
- \exp \left(-(n+n')^2 \over L \right) \right].
\label{oo}
\end{equation}
Hence for 
\begin{equation}
\label{scaling}
i_p = Lx_p \ \ \ \ , \ \ \ \  n_{i_p}= L^{1/2} y_p \ \ \ ,
\end{equation}
one gets
$$
\langle n_{i_p} | (D+E)^{i_{p+1}-i_{p}} | n_{i_{p+1}} \rangle
 \simeq  {4^{L (x_{p+1}-x_p)} \over \sqrt{L}} \
g(y_p,y_{p+1};x_{p+1}-x_{p})
$$
where $g$ is defined as in (\ref{gdef}).
This formula remains valid even if $n_{i_p}$ and/or $n_{i_{p+1}}$ are of order
$1$. {}For example one obtains that way
that if $n_{i_p}$ is of order $1$ and  $n_{i_{p+1}} \simeq \sqrt{L} y_{p+1}$,
$$
\langle n_{i_p} | (D+E)^{i_{p+1}-i_p} | n_{i_{p+1}} \rangle
 \simeq  {4^{L(x_{p+1}-x_p)} \over  \sqrt{\pi} L  }  \  2 n_{i_{p}}  \
h(y_{p+1};x_{p+1}-x_p)
$$
so that  (\ref{sumw}) and(\ref{oo})  give
 \begin{equation}
 \langle W | (D+E)^L | V \rangle  \simeq  {4^{L+1} \over \sqrt{\pi} L^{3/2}}
\ {(1- \alpha) \alpha \over (2 \alpha -1)^2 }
\ {(1- \beta ) \beta  \over (2 \beta  -1)^2 } .
\label{denom-asympt}
 \end{equation}

The correlation function $\overline{n_{i_1} ... n_{i_k}}$ of the heights of
a walk $w$ at positions $i_1,... i_k$ is then given by
\begin{eqnarray*} 
\overline{n_{i_1} ... n_{i_k}} &=& \sum_{n_0,n_{i_1},... n_{i_k}, n_L}
n_{i_1}... n_{i_k} \times \\
&&  {\langle n_0 | (D+E)^{i_1} | n_{i_1} \rangle .... 
\langle n_{i_k} | (D+E)^{L-i_k} | n_L \rangle
 \left( 1 - \alpha \over \alpha \right)^{n_0}
 \left( 1 - \beta \over \beta \right)^{n_L}
\over \sum_w \Omega_{w}}, 
\end{eqnarray*}
and  for $i_1 = L x_1,..., i_k = L x_k$ one gets
$$\overline{n_{i_1} ... n_{i_k}} = L^{k/2}  \ \overline{Y(x_1)...Y(x_k)} $$
which  using (\ref{eq2}) leads to  (\ref{cor}).

\subsection{Derivation of (\ref{Qmu})}
{}For large $L$, with $n, n'$ and $N- L/2$ of order $\sqrt{L}$,
one can easily see from
(\ref{XLNexp}) that
$$\langle n | X_{L}(N) | n' \rangle \simeq 2 { 4^L \over \pi L} \left[
e^{-2 {(\Delta N)^2 \over L} - 2 { ( \Delta N + n-n')^2 \over L} }
-e^{-2 {(\Delta N +n )^2 \over L} - 2 { ( \Delta N -n')^2 \over L} }
\right], $$
where $\Delta N = N- L/2, $
which can be rewritten as
$$\langle n | X_{L}(N) | n' \rangle \simeq 2 \  { 4^L \over \pi L} e^{-
(  2 \Delta N +n- n')^2 \over L} \left[
e^{- {(n  -n')^2 \over L}} -  
e^{- {(n  +n')^2 \over L}  } 
\right]. $$
Thus 
$F_{L,N}(n,n')$  defined in (\ref{F}) becomes for $n,n'$ and $\Delta
N$ of order $\sqrt{L}$ 
$$F_{L,N}(n,n') \simeq { 2 \over \sqrt{\pi L}} \exp \left[ - (2  \  \Delta N
+ n- n' )^2 \over L \right], $$
 and this shows that in the large $L$ limit  (\ref{qL}) reduces to (\ref{Qmu}).

\subsection {Origin of (\ref{BM+BE})}
As already noted in section 1, expression (\ref{BM+BE}) is essentially
equivalent 
to (\ref{Qmu}) so the derivation of (\ref{Qmu}) above also gives
(\ref{BM+BE}).  It is interesting, however, to understand the origin
of (\ref{BM+BE}) directly from the microscopic picture involving the
walks $w$.  To do that let us define the joint distribution  $\tilde
\nu(w,\tau)$ of $w$
and $\tau = \{ \tau_i \}$.  It follows directly from
(\ref{eq1})
 that
\begin{equation}
\tilde \nu(w,\tau) = \nu(w) P(\tau|w),
\label{2.29}
\end{equation}
where  $\nu(w)$ is defined in (\ref{nu})
and the conditional probability of the $\{ \tau_i \}$ given $w$ is a
product measure
\begin{equation}
P(\tau|w) = \prod_{i=1}^L {1 + (n_i - n_{i-1})(2 \tau_i - 1) \over 2
}. 
\label{2.31}
\end{equation}

This leads to a simple way of generating steady state configurations
of the occupation numbers $\{ \tau_i\}$. 
First one generates a random walk $w$ of $L$ steps according to the
measure $\nu(w)$. Then according to (\ref{2.31}) 
a steady state configuration  $\{ \tau_i\}$ is obtained by taking
$\tau_i=1$ whenever $n_i(w)-n_{i-1}(w)=1$, 
$\tau_i=0$ whenever $n_i(w)-n_{i-1}(w)=-1$ and by choosing $\tau_i=0$ or
$1$  with equal probabilities for each $i$ such that  $n_i(w)-n_{i-1}(w)=0$.
Therefore the fluctuations of the density have two contributions: the
random choice of the walk $w$ which is at the origin of  
$\dot Y(x)$ in (\ref{BM+BE}) and once $w$ is chosen, the random
choices of the $\tau_i$ for the flat parts of the walk, which
correspond to  
$\dot B(x)$  in (\ref{BM+BE}).  The normalization of $B(x)$ in
(\ref{1.9}) arises from the fact that, for large $L$, $n_i(w) =
n_{i-1}(w)$ for approximately half the steps of the walk
$w$.  

\section{Simulations}
An interesting question is to know whether the fluctuations of density
of  one dimensional driven diffusive systems in their 
maximal current phase take always the form (\ref{BM+BE}), once
properly normalized. 

First, we believe that our results also hold for the general ASEP where
particles can also jump to the left with rate $q < 1$.
In this case also, one can choose a representation of the matrices,
used recently to calculate the large deviation function in the weak
asymmetry limit \cite{ED}, 
such that matrix elements can be thought as sums over weighted walks
which do not cross the origin.  We will not discuss this further here.

In order to test whether the results obtained here for the TASEP remain valid
for  a broader class of  models, we
consider in this section a generalisation of the TASEP 
   \cite{LC,SZL} in which particles   are  extended.
In this model, each particle occupies $d$ consecutive sites
of a lattice of $L+d-1$ sites and the exclusion rule forbids that any site 
of the lattice is occupied by more than one particle.  There are thus 
$L$ possible positions for a single particle of size $d$ on the
lattice of $L + d - 1$ sites.  
By convention, we say that a particle is at site $i$ when 
it covers sites $i,i+1,...i+d-1$.  
The system evolves according to the following rule:
during each infinitesimal time interval $dt$, each particle jumps one
step  to its
right with probability $dt$ provided that this is allowed by the 
exclusion rule. Moreover if the first $d$ sites  are empty,
a new particle is injected at site $1$ with probability $ \alpha dt$ and
if a particle covers sites $L,L+1,... L+d-1$, it is removed with
probability $ \beta dt$.
{}For $d=1$ the problem reduces to the TASEP discussed in sections 1 and 2.

{}For general $d$, the current $J_d$ and the density  $\rho_d$ in the
maximal current phase    are given by \cite{LC}
\begin{equation} J_d= {1 \over (\sqrt{d} +1)^2}
\label{Jd}
\end{equation}
\begin{equation}
\rho_d= {1 \over \sqrt{d} (\sqrt{d} +1)}
\label{rhod}
\end{equation}
These expressions can be understood by considering a system of $M$ such
particles on a ring of $L$ sites, and by using the fact that all
allowed configurations 
are equally likely (see the appendix).

In this ring geometry,  the number $m$ of particles on $l$ consecutive
sites fluctuates in the steady state and it is possible to show 
(see the appendix)
that for $d   \ll l \ll L$, one has
\begin{equation}
\label{dm}
{\langle m^2 \rangle - \langle m \rangle^2  \over l}\simeq
\Delta_d(\rho)  
\end{equation}
 with $$\Delta_d(\rho)= \rho(1-\rho) - \rho^2 (d-1) (2 - d \rho) ,
$$
This expression can also be recovered from the pressure \cite{LY} of
 this system 
 $p= \log\left(\frac{1-\rho (d-1)}{1-\rho d}\right)$ 
 using $\Delta_d(\rho)=\rho {d \rho \over dp}$.
%where $p$ is the pressure of this system \cite{LY}.
In our simulations we tried to see whether, for large system sizes with
open boundary conditions, the
fluctuations of density in the maximal current phase would still be
given by the statistics of a 
Brownian excursion with (\ref{BM+BE}) and (\ref{rxxp}) being replaced by
\begin{equation}
 \rho(x) - \rho_d   = \sqrt{ \Delta_d(\rho_d) \over L} \ [\dot B(x) +
\dot Y(x) ]
\label{BM+BEd}
\end{equation}
and
\begin{equation}
 r(x,x') = \int_x^{x'} dy \left(\rho(y) - \rho_d \right) = \sqrt{
\Delta_d(\rho_d) \over L}
 [B(x') + Y(x') - B(x) - Y(x)]  
\label{rxxpd}
\end{equation}

If we look at the statistical properties of the normalized number $s(x)$
of particles in a box of size $Lx$ centered at
the middle of the system
\begin{equation}
s(x) = \sqrt{\frac{L}{\Delta_d(\rho_d)}} \ r\left({1-x\over 2}, {1+x\over
2} \right)
\end{equation}
we get, using (\ref{rxxpd}) and the properties (\ref{pro}) of $Y(x)$
\begin{equation} \label{s2th}
\overline{s(x)^2} = {  2 x - 3 x^2  \over 2 } +
 {3 - 2 x + 3 x^2  \over  2 \pi   }   \cos^{-1}\left( {1-x\over 1+x }
\right)
-{ 3 (1-x) \sqrt{x} \over  \pi  }
\end{equation}
and
\begin{eqnarray} \label{s4th}
\overline{s(x)^4} &=& \frac{3 x^2  (x-2 ) (5x-2 )}{4}
   + \frac{ 15- 3x^2 (x-2 ) ( 5 x -2)}{4 \pi} \cos^{-1}\left( {1-x\over 1+x }
\right) \nonumber \\&& +\frac{1-x}{ 2 \pi (1+x)} \sqrt{x} (15 x^3-11
   x^2-25 x -15) 
\end{eqnarray}
\begin{figure}%[ht]
\begin{center}
\begin{picture}(0,0)%
\includegraphics{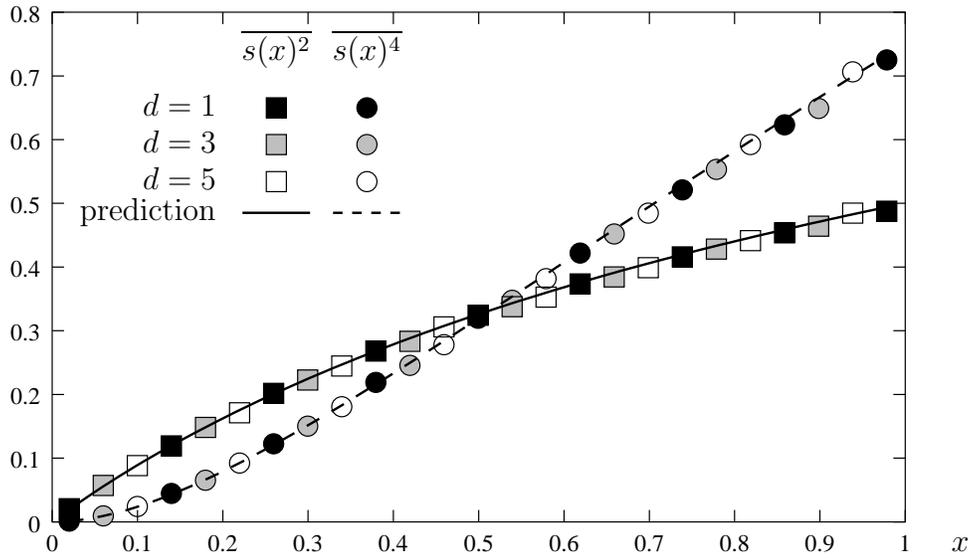}%
\end{picture}%
\setlength{\unitlength}{3947sp}%
\begingroup\makeatletter\ifx\SetFigFont\undefined%
\gdef\SetFigFont#1#2#3#4#5{%
  \reset@font\fontsize{#1}{#2pt}%
  \fontfamily{#3}\fontseries{#4}\fontshape{#5}%
  \selectfont}%
\fi\endgroup%
\begin{picture}(5962,3478)(474,-3044)
\put(1762,-258){\makebox(0,0)[rb]{\smash{\SetFigFont{12}{14.4}{\familydefault}{\mddefault}{\updefault}$d=1$}}}  
\put(1762,-489){\makebox(0,0)[rb]{\smash{\SetFigFont{12}{14.4}{\familydefault}{\mddefault}{\updefault}$d=3$}}}  
\put(1762,-722){\makebox(0,0)[rb]{\smash{\SetFigFont{12}{14.4}{\familydefault}{\mddefault}{\updefault}$d=5$}}}
\put(2710,
87){\makebox(0,0)[b]{\smash{\SetFigFont{12}{14.4}{\familydefault}{\mddefault}{\updefault}$\overline{s(x)^4}$}}}
\put(2141,
87){\makebox(0,0)[b]{\smash{\SetFigFont{12}{14.4}{\familydefault}{\mddefault}{\updefault}$\overline{s(x)^2}$}}}
\put(1762,-915){\makebox(0,0)[rb]{\smash{\SetFigFont{12}{14.4}{\familydefault}{\mddefault}{\updefault}prediction}}}
\put(6436,-2999){\makebox(0,0)[b]{\smash{\SetFigFont{12}{14.4}{\familydefault}{\mddefault}{\updefault}$x$}}}
\end{picture}
\end{center}
\caption{The moments $\overline{s(x)^2}$ and
$\overline{s(x)^4}$ of the fluctuations of the number of particles $s(x)$
(normalized as in  (\ref{sNx})) between sites $L \frac{1-x}{2}$ and $L
\frac{1+x}{2}$ for a system of $L=601$ sites and for 3 sizes of particles
(d=1, 3, 5). The curves are the analytic predictions (\ref{s2th}) and
(\ref{s4th}).} \label{figure}
\end{figure}

We have simulated this model for  three sizes of particles $d=1$, $3$
and $5$ for an open system of  
$L=601$  sites when $\alpha=\beta=1$. 
Let $N_x$ be the number of particles in the interval  
$L \frac{1-x}{2} \leq i \leq L \frac{1+x}{2}$ (we count as $N_x$ the
number of sites in the interval covered by a particle, divided by the
length $d$ of one 
particle).
 The normalized number of
particles $s(x)$ is related to $N_x$ by
\begin{equation}
s(x)=\frac{N_x-L x \rho_d} {\sqrt{L \Delta_d(\rho_d)}}
\label{sNx}
\end{equation}
 We have measured the second
and fourth moments of $s(x)$  in the steady state, averaged over typically
$10^8$
updates per site.
In figure \ref{figure}, we compare the results of our simulations with
the theoretical predictions 
(\ref{s2th}) and (\ref{s4th}). 
The fact that the curves for the different choices of $d$ coincide
indicate that the
fluctuations, once properly normalised, are universal (i.e. do not depend
on $d$)

\section{Conclusion}

The main result of the present paper is that the steady state
fluctuations of the density profile in the TASEP can be written in terms
of
a Brownian excursion as in (\ref{cor},\ref{Qmu},\ref{BM+BE}).

Our simulations of a more general model in which particles are
extended indicates that the fluctuations of the density profile may be
universal. It would of course be nice to make other numerical tests
of this universality and to see whether it could be understood by a 
more macroscopic approach, such as in 
\cite{BDGJL,BDGJL2}.

Another interesting question would be to see whether the time dependent
fluctuations of the density profile would arise from some simple
stochastic dynamics of the Brownian excursion.

\section*{Appendix}
Let us consider the TASEP
 on a ring of L sites 
with $N$ particles of size  $d$ as in section 3. This hard rod problem
has a long history starting with the works of Lee and Yang \cite{LY}. 
All allowed configurations (i.e. configurations  which satisfy the
 exclusion rule) are equally likely 
and let us call  $Z_L(N)$  their number.

The number $ z_L(N)$ of configurations of $N$ particles on a lattice of
$L$ sites with open boundary conditions is given by, 
 $z_0(N)=z_1(N)=z_2(N)=... z_{d-1}(N)= \delta_{N,0}$, and for $L > Nd$
$$z_L(N) = {(L-Nd + N)! \over N! \ (L-Nd)!} $$
(one can easily check that $z_L(N)$ satisfies the recursion 
$z_L(N)= z_{L-1}(N)+z_{L-d} (N-1)$).

By considering that site $1$ on a ring is either empty or covered by one
particle, one can 
express $Z_L(N)$ in terms of the partition functions of the open systems
$$ Z_L(N) = z_{L-1} (N) + d z_{L-d}(N-1)  
 = { L \ (L-Nd + N-1) !  \over N! \ (L-Nd)! }$$
The current $J$ on the ring is given by
$$J= {z_{L-d-1}(N-1) \over Z_L(N)} = { N (L - Nd) \over L ( L - Nd + N
-1) } $$
{}For large $L$ and $N$, at fixed density $\rho=N/L$, $J$ becomes
$$J= {\rho (1- \rho d )\over 1 - \rho(d-1) } $$
which gives the expressions (\ref{rhod}) and (\ref{Jd}) when the current
is maximal.

One can calculate the correlation function $\langle \tau_i \tau_{i+k}
\rangle $ between the occupations on  the ring
$$ \langle \tau_i \tau_{i+k} \rangle = {1 \over Z_L(N)} \sum_{n =0  }^N
z_{k-d }(n) \  z_{L-k-d}(N-n-2) $$
Using the above expressions of $Z_L(N)$ and $z_L(N)$, one gets for large
$L$ and $N$, keeping $\rho=N/L, k$ and $n$ fixed
$$ \langle \tau_i \tau_{i+k} \rangle =  \sum_{n \geq 0}
z_{k-d } (n)  \ 
 \rho^{n+2} \ (1-\rho d)^{k - nd - d} \  [1-\rho(d-1)]^{(n+1)(d-1)-k}
$$
then  as 
$$ \sum_{L \geq Nd} z_L(N) x^L = { x^{Nd} \over (1-x)^{N+1}}$$
one gets that (for arbitrary $\epsilon$)
$$\sum_{k \geq 1}\langle \tau_i \tau_{i+k} \rangle e^{-k \epsilon} =
{\rho^2 e^{- d \epsilon} \over (1 - \rho d)(1 - e^{-\epsilon}) +  \rho (1
- e^{- d \epsilon})}$$
and as
$$\sum_{k \geq 1}\langle \tau_i \rangle \langle\tau_{i+k} \rangle e^{-k
\epsilon} = {\rho^2 e^{-\epsilon} \over 1 - e^{-\epsilon}} $$
one finds by expanding the above expressions in powers of $\epsilon$
that
$$\sum_{k \geq 1}\langle \tau_i \tau_{i+k} \rangle -\langle \tau_i
\rangle \langle\tau_{i+k} \rangle = -{(d-1) \rho^2 (2 -  \rho d) \over 2}
$$
so that the fluctuations of the number $m$ of particles on $l$ consecutive
sites, with $d \ll l \ll L$
is given by
$${\langle m^2 \rangle - \langle m \rangle^2 \over l} = (\rho-
\rho^2) - (d-1) \rho^2 (2 - \rho d) $$
as given in (\ref{dm}).

\section*{Acknowledgments}

We would like to thank our collaborator, E. Speer, who made very
substantial contributions to this work.
B.D. and   J.L.L.  acknowledge the hospitality of the
 Institute for
Advanced Study, Princeton,  where   this work was started.   The work
of J.L.L. was supported by NSF Grant MR 01-279-26, AFOSR Grant AF
49620-01-1-0154, and 
DIMACS and its supporting agencies, the NSF under
contract STC-91-19999 and the N.  J.  Commission on Science and
Technology.

\section*{Figure Captions}

Figure \ref{figure}: The moments $\overline{s(x)^2}$ and
$\overline{s(x)^4}$ of the fluctuations of the number of particles $s(x)$
(normalized as in  (\ref{sNx})) between sites $L \frac{1-x}{2}$ and $L
\frac{1+x}{2}$ for a system of $L=601$ sites and for 3 sizes of particles
(d=1, 3, 5). The curves are the analytic predictions (\ref{s2th}) and
(\ref{s4th}).

\end{document}